\documentstyle[12pt]{article}
\newtheorem{theorem}{Theorem}
\newtheorem{itlemma}{Lemma}[section]
\newtheorem{itproposition}[itlemma]{Proposition}
\newtheorem{itcorollary}[itlemma]{Corollary}
\newtheorem{itremark}[itlemma]{Remark}
\newtheorem{itremarks}[itlemma]{Remarks}
\newtheorem{itdefinition}[itlemma]{Definition}
\newtheorem{itexample}[itlemma]{Example}

\newenvironment{lemma}{\begin{itlemma}\rm}{\end{itlemma}} %no-italics
\newenvironment{remark}{\begin{itremark}\rm}{\end{itremark}} %no-italics
\newenvironment{remarks}{\begin{itremarks} \rm}{\end{itremarks}}
\newenvironment{corollary}{\begin{itcorollary}\rm}{\end{itcorollary}}
\newenvironment{proposition}{\begin{itproposition}\rm}{\end{itproposition}}
\newenvironment{definition}{\begin{itdefinition}\rm}{\end{itdefinition}}
\newenvironment{example}{\begin{itexample}\rm}{\end{itexample}}
\newenvironment{fact}{\noindent {\em Fact}. \ \ }{\hfill \medskip}
\newenvironment{proof}{\noindent {\em Proof}.\ \
}{\hspace*{\fill}$\Box$\medskip}
\newenvironment{claim}{\noindent {\em Claim}. \ \ }{\hfill \medskip}
\newcommand{\be}[1]{\begin{equation}\label{#1}}
\newcommand{\ee}{\end{equation}}
\newcommand{\bl}[1]{\begin{lemma}\label{#1}}
\newcommand{\br}[1]{\begin{remark}\label{#1}}
\newcommand{\brs}[1]{\begin{remarks}\label{#1}}
\newcommand{\bt}[1]{\begin{theorem}\label{#1}}
\newcommand{\bd}[1]{\begin{definition}\label{#1}}
\newcommand{\bp}[1]{\begin{proposition}\label{#1}}
\newcommand{\bc}[1]{\begin{corollary}\label{#1}}
\newcommand{\bfact}[1]{\begin{fact}\label{#1}}
\newcommand{\bex}[1]{\begin{example}\label{#1}}
\newcommand{\ec}{\end{corollary}}
\newcommand{\efact}{\end{fact}}
\newcommand{\eex}{\end{example}}
\newcommand{\el}{\end{lemma}}
\newcommand{\er}{\end{remark}}
\newcommand{\ers}{\end{remarks}}
\newcommand{\et}{\end{theorem}}
\newcommand{\ed}{\end{definition}}
\newcommand{\ep}{\end{proposition}}
\newcommand{\epr}{\end{proof}}
\newcommand{\bpr}{\begin{proof}}
\newcommand{\bcl}{\begin{claim}}
\newcommand{\ecl}{\end{claim}}

\newcommand{\bi}{\begin{itemize}}
\newcommand{\ei}{\end{itemize}}
\newcommand{\ben}{\begin{enumerate}}
\newcommand{\een}{\end{enumerate}}
\newcommand{\text}[1]{\hbox{\rm \ #1\ \/}}

\newcommand{\CC}{\mbox{${\rm \:  C\!\!\! I
\;\;}$}}

\newcommand{\RR}{\mbox{${\rm \:  R\!\!\!\! I
\;\;}$}}

\newcommand{\vs}{\vspace{0.25cm}}

\newcommand{\rhof}{{{\cal F}_{\rho}}}
\newcommand{\tildec}{{\tilde{\cal C}}}
\newcommand{\m}{{{\cal M}}}
\newcommand{\n}{{{\cal N}}}

\newcommand{\q}{{{\cal Q}}}
\newcommand{\r}{{{\cal R}}}

\begin{document}

\noindent \hfill 
\begin{center}
{\Large {The Lie Algebra Structure and Nonlinear 
Controllability of Spin Systems}}
\end{center}

\bigskip

\begin{center}

{Francesca Albertini\\ \vs Dipartimento di Matematica Pura ed
Applicata,\\ Universit\`a di Padova,\\ via Belzoni 7,\\ 35100
Padova, Italy.\\ Tel. (+39) 049 827 5966\\ email:
albertin@math.unipd.it}\\

\vs \vs {Domenico D'Alessandro \\ \vs Department of Mathematics\\
Iowa State University \\ Ames, IA 50011,  USA\\ Tel. (+1) 515 294
8130\\ email: daless@iastate.edu}

\end{center}

\vspace{0.5cm}

\begin{abstract}

In this paper, we study the controllability properties 
and the Lie algebra structure of networks of particles 
with spin immersed in an electro-magnetic field. 
We relate the Lie algebra structure to the properties 
of a graph whose nodes represent the
particles and an edge connects two nodes if and only if  the
interaction between the two 
corresponding particles is active. For networks with different
gyromagnetic ratios,  we  provide a
necessary and sufficient condition of  controllability in terms of
the properties of the above mentioned  graph and describe the Lie algebra
structure in every case. For these systems all the controllability
notions, including the possibility of driving the evolution operator
and/or the state, are equivalent. For general networks (with possibly
equal gyromagnetic ratios), we give a
sufficient condition of controllability.  A general form of interaction among 
the particles is assumed which includes both Ising and Heisenberg
models as special cases. 

Assuming Heisenberg interaction we provide an analysis
of low 
dimensional cases (number of particles less then or equal to three)
which include necessary and sufficient controllability conditions as
well as a study of their Lie algebra structure. 
This also, provides an example of quantum mechanical systems where 
controllability of the state is verified while controllability of the 
evolution operator is not. 
\end{abstract}



\section{Introduction}

The controllability of multilevel quantum mechanical
 systems described by bilinear
models can be investigated using
results on the controllability of bilinear
systems varying on Lie
groups \cite{Tarn}, \cite{murti}. In 
particular, general results established 
in \cite{SUSS} can be applied to this case leading to 
the calculation of the Lie algebra generated by the
Hamiltonian of the system and 
the verification of a rank condition. The determination of 
this Lie algebra for {\it classes}
of quantum systems is a problem of both 
fundamental and practical importance in the theory of quantum
control. In fact, it gives the set of states that can be obtained by
driving the system opportunely and letting it evolve for an appropriate
amount of time. Previous work in this direction, for various classes of
quantum systems, was done in \cite{S1}, \cite{S2}.

In this paper, we
analyze the Lie algebra structure and give conditions of controllability
for  a network  of 
interacting spin
$\frac{1}{2}$ particles in a
driving  electro-magnetic field. Spin $\frac{1}{2}$
particles are of great interest because they can
be used as elementary pieces of information (quantum
bits) in quantum information theory \cite{divi}. These systems  
can be driven with techniques of Nuclear Magnetic Resonance  
\cite{QCNMR}. A study of
their  controllability properties   gives 
information on what state transfers can be obtained
with a given physical set-up. A previous study on the controllability of
this system was carried out in \cite{KG}, \cite{thomas}. Results on the
controllability of  systems of one and two spin $\frac{1}{2}$
particles can be found in \cite{ioSCL}, \cite{BKS}.

%%%%%%%%%%%%%%%%%%%%%%%%%%%%%%%%%%%%%%%%%%%%%%%%%%%%%%%%%%%%%%
In the present  paper we relate the 
Lie algebra structure of a network of spin
$\frac{1}{2}$ particles to the properties of a graph whose nodes
represent the particles and whose edges represent the interaction
between the particles. We analyze first 
the case of networks with particles with different
gyromagnetic ratios. For these systems we give a necessary and
sufficient condition of controllability in terms of connectedness of
the associated  graph and describe the 
Lie algebra structure in every case. It will 
follow from this analysis that all the controllability
conditions are equivalent for this class of systems. In particular it is
possible to drive the {\it state} of the  system to any
configuration if and only if it is possible to drive the {\it evolution
operator} to any unitary operator. We consider then systems with possibly
equal gyromagnetic ratio and give a sufficient condition of
controllability in this case. Complete results including necessary and
sufficient conditions of various types of controllability are obtained for low
dimensional cases, namely for a number of particles $\leq 3 $. These cases
are the most common in  practical applications. We assume here (for the
case number of particles $=3$) an Heisenberg model for the interaction
between particles.  In this analysis we
also display an example of a model which is controllable in the state
but not controllable in the evolution operator.

The paper is organized as follows. In Section \ref{cqms} we review general
notions of controllability for quantum mechanical systems. We 
recall some results proved in \cite{noi} about the relation 
among different notions of controllability as well as some of the results of 
\cite{Tarn}, \cite{SUSS}, \cite{murti} about 
controllability of quantum systems. 
In Section \ref{misp}, we
describe the general model of systems of $n$ interacting spin
$\frac{1}{2}$ particles
 and define some
notations used in the paper. In Section \ref{CCS} we prove a Lemma which
describes a particular subalgebra of the total Lie algebra, that we call
the `Control subalgebra'. This will play an important role in the
following development. In Section \ref{las} we study  the Lie algebra
structure associated to the model described in Section \ref{misp}
assuming that all the particles have different gyromagnetic ratios. In
Section \ref{SPEGR},  we remove  this assumption and prove a general sufficient
condition of controllability. We study low dimensional cases in Section
\ref{lds} and give some conclusions in Section \ref{c}.

\section{Controllability of Quantum Mechanical Systems}
\label{cqms}

In many physical situations 
the dynamics of a multilevel quantum
system can be described by Schr\"odinger equation in the form, 
\cite{conmoh}, \cite{murti}, \be{gensys} \dot
{|\psi>}=H{|\psi>}=(A+\sum_{i=1}^mB_iu_i(t)){|\psi>}, \ee where
$|\psi>$\footnote{We use Dirac notation  $|\psi>$ to denote a
vector on $\CC^n$ of length $1$, and  $<\psi|:=|\psi>^*$ where
$^*$ denotes transposed conjugate.} is the state  vector varying
on the  complex sphere $S^{n-1}_{\CC}$ defined as the set of
$n$-ples of complex numbers $x_j+iy_j$, $j=1,...,n$,  with
$\sum_{j=1}^n x_j^2+y_j^2 =1$. $H$ is called the Hamiltonian 
of the system. The matrices $A,$ $B_1,...,B_m$ are
in the Lie algebra of {\it skew-Hermitian} matrices of dimension
$n$, $u(n)$. If $A$ and $B_i$, $i=1,...,m$, have zero trace, they
are in the Lie algebra of skew Hermitian matrices with zero trace
$su(n)$\footnote{Since trace of $A$ and $B_i$, $i=1,2,...,m$, only
introduce a phase factor in the solution of (\ref{gensys}), and
states that differ by a phase factor are physically
indistinguishable, it is possible to transform the equation
(\ref{gensys}) into an equivalent one of the same form where the
matrices $A$ and $B_i$, $i=1,...,m$, are skew-Hermitian and with
zero trace, namely they are in $su(n)$.}. The
functions $u_i(t)$, $i=1,2,...,m$, are time varying 
components of electro-magnetic fields that play the 
role of {\it controls}. They are
 assumed to be piecewise continuous,
however the considerations in the
following would not change had we
considered other classes of controls
such as piecewise constant
  or bang bang controls.

The solution of (\ref{gensys}) at time $t$, $|\psi(t)>$ with
initial condition $|\psi_0>$ is given by \be{solutione}
|\psi(t)>=X(t)|\psi_0>, \ee where $X(t)$ is the solution at time
$t$ of the equation \be{jasbd} \dot
X(t)=(A+\sum_{i=1}^mB_iu_i(t))X(t), \ee with initial condition
$X(0)=I_{n \times n}$. The solution $X(t)$ varies on the Lie group
of unitary matrices $U(n)$ or the Lie group of special unitary
matrices $SU(n)$ if the matrices $A$ and
$B_i$ in (\ref{jasbd}) have zero trace.

\vs

Various notions of controllability can be defined for system
(\ref{gensys}). In particular, we will consider the following
three. 

\begin{itemize}
\item System (\ref{gensys}) is said to be
{\it Operator Controllable} if it is possible to drive $X$ in
(\ref{jasbd}) to any value in $U(n)$ (or $SU(n)$). 
\item System (\ref{gensys}) is {\it State Controllable} if it is
possible to drive the state 
$|\psi>$ to any value on the complex
sphere $S_{\cal C}^{n-1}$, for any given 
initial condition. 
\item System (\ref{gensys}) is said
to be  {\it Equivalent State Controllable} if it is possible to
drive the state $|\psi>$ to any value on the complex sphere modulo
a phase factor $e^{i \phi}$, $\phi \in \RR$. 
\end{itemize}
{}From a physics point
of view, equivalent state controllability is equivalent to state
controllability since states that differ only by a phase factor are
physically indistinguishable.

{}From the expression (\ref{solutione}) for $|\psi>$, it is clear
that state controllability is related to the possibility of
driving $X$ to a subset of $SU(n)$ or $U(n)$ which is transitive
on the complex sphere. Transitivity of transformation groups on
spheres was studied in  
\cite{Borel}, \cite{Samelson1}, \cite{MZ},  
\cite{Samelson2} and the 
necessary connections for application to
quantum mechanical systems 
where made in \cite{noi}. In the
following theorem, we summarize some 
of the  results of  \cite{noi} that will be used in the following. 
Part 2) of the Theorem was proved in   \cite{Tarn},
\cite{SUSS}, \cite{murti}. 
Here and in the 
following we will denote 
by $\cal L$ the Lie algebra generated  by 
$A,B_{1},\ldots,B_{m}$ in (\ref{gensys}). \bt{sommarize} \hfill
\begin{enumerate}
\item
A quantum mechanical system (\ref{gensys}) is state controllable
if and only if it is equivalent state controllable. Both these
conditions are implied by operator controllability.
\item
The system is operator controllable if and only if the Lie algebra
$\cal L$ generated by the matrices $A,B_1,....,B_m$ is $u(n)$ or
$su(n)$.
\item
The system is state controllable if and  only if 
${\cal L}$ is $su(n)$ or $u(n)$, or, in the case of $n$ even, 
isomorphic to $sp(\frac{n}{2})$\footnote{Recall the Lie algebra
 of symplectic matrices $sp(k)$ is
the Lie algebra  of matrices $X$
in $su(2k)$ satisfying $XJ+JX^T=0$, with
$J$ given by $J=\pmatrix{0 & I_{k \times k} \cr -I_{k \times k} &0}$}.
\item Consider the $n \times n$ matrix with $i$ in the position 
$(1,1)$ and zero everywhere else. Call this matrix $D$. Let 
$\cal D$ be the subalgebra of $\cal L$ of matrices that commute with 
$D$. Then, the system is state controllable if and only if 
$\dim {\cal L} - \dim {\cal D}=2n-2$. 
\item Assume $n$ even. There is no subalgebra of $su(n)$ which 
contains properly any subalgebra isomorphic to 
$sp(\frac{n}{2})$ other than $su(n)$ itself.
\end{enumerate}
\et
Because of the equivalence between state controllability and 
equivalent state controllability, in the sequel we will only refer to the 
two notions of state controllability and operator controllability. Controllability notions in a
density matrix description of quantum 
dynamics were considered  in \cite{noi}.  

\section{Model of interacting spin $\frac{1}{2}$ particles}
\label{misp}

From this point on, we will denote by $n$ (which in the previous 
section denoted the dimension of a general quantum system) the number 
of spin $\frac{1}{2}$ particles in a network. 
The state dimension of this system 
is $2^{n}$.

To define the model we will study, we first need to
recall some definitions. The following three 
matrices in $su(2)$ are
called  {\it {Pauli
matrices}} (see e.g. \cite{sakurai}):
\be{Pauli}
\sigma_x:= \frac{1}{2} \pmatrix{0 & 1 \cr 1 & 0}, 
\qquad \sigma_y:=\frac{1}{2} \pmatrix{0 & -i \cr
i & 0},  \qquad \sigma_z:=\frac{1}{2}\pmatrix{1 & 0 \cr
0 & -1}.
\ee
The Pauli matrices satisfy the fundamental commutation relations
\be{commu}
[\sigma_x,\sigma_y]=i\sigma_z ; \qquad [\sigma_y,\sigma_z]=i\sigma_x; \qquad
[\sigma_z,\sigma_x]=i\sigma_y.
\ee
It is known that the matrices 
$i \sigma_x,$ $i\sigma_y$, $i\sigma_z$ form a
basis in $su(2)$.  Moreover, the
 set of matrices
$i(\sigma_1 \otimes \sigma_2 \otimes 
\cdot \cdot \cdot \otimes \sigma_n)$, where
$\sigma_j$, $j=1,...n$, is equal to one of the
Pauli matrices or the $2 \times 2$ identity
$I_{2\times 2}$,  without
 $i(I_{2\times 2} \otimes I_{2\times 2}\otimes \cdot \cdot \cdot
\otimes I_{2\times 2})$, form a basis in $su(2^n)$.
(Here $\otimes$ indicates the Kronecker  product for matrices.)

In the following, we will use the notation $I_{kx}$ for the
Kronecker  product \be{Kronecker} I_{kx}:=\sigma_1 \otimes
\sigma_2 \otimes \cdot \cdot \cdot \otimes \sigma_n, \ee where all
the the elements $\sigma_j$, $j=1,...,n$ are equal to the $2\times
2$ identity matrix, except the $k-$th element which is equal to
$\sigma_x$. More in general, we will use the notation $I_{k_1 l_1,
k_2 l_2,...,k_rl_r}$, with $1\leq k_1 < k_2< \cdot \cdot \cdot <
k_r\leq n$ and $l_j=x,y$ or $z$, $j=1,...,r$, for a Kronecker
product of the form (\ref{Kronecker}) where all the $\sigma_j$ are
equal to the identity $I_{2 \times 2}$ except the ones in the
$k_j-$th positions which are equal to the Pauli matrices
$\sigma_{l_j}$. The matrices so defined (excluding the identity
matrix) span $su(2^n)$.  Some elementary properties of the
commutators of the matrices just defined that will be used in the
following are collected in Appendix $A$.

The Hamiltonian of the system of $n$ interacting spin
$\frac{1}{2}$ particles in a driving electro-magnetic
field is given in the form \cite{BrockKhaneja}:
\be{formaHamiltoniano} H=H_0+H_I. \ee
Here $H_0$, which denotes 
the {\it internal (or unperturbed) Hamiltonian}, 
 is given by
\be{internal} H_0:=\sum_{k<l}^n  (M_{kl}I_{kx,lx}+ N_{kl}I_{ky,ly}+
P_{kl}I_{kz,lz}), \ee where ${M_{kl},N_{kl},P_{kl}}$ are 
the coupling constants between
particle $k$ and particle $l$. This general 
model of the interaction
between different particles includes as special cases both the Ising and
the Heisenberg model (\cite{Mahan}, pg. 46). 
The term  $H_I$, {\it Control   
Hamiltonian},  is given by
\be{interactionHamiltonian}
H_I:=(\sum_{k=1}^n \gamma_k I_{kx})u_x(t)+ (\sum_{k=1}^n \gamma_k
I_{ky})u_y(t)+ (\sum_{k=1}^n \gamma_k I_{kz})u_z(t),
\ee
where $u_x$, $u_y$ and $u_z$ are the
$x$, $y$ and $z$ components of the
electro-magnetic field and $\gamma_j$, $j=1,2,...,$ is the
gyromagnetic ratio of the $j$-th particle. In general, we 
assume that we are able to vary all the three components
of the magnetic field for control (cfr. Remark
\ref{RDM}). 
%In practical laboratory set-ups, an ensemble 
%of identical systems is controlled and 
%the $z-$component of the control is held 
%constant. Also, different forms for the interaction 
%(\ref{internal}) can be considered. As a consequence, a different 
%model has to be considered for practical applications. However, this 
%does not modify the main results given in the sequel  as we discuss in 
%Remark \ref{RDM} below. 
Schr\"odinger equation (\ref{jasbd})
for the evolution  matrix $X$ has the form,  
\be{scrod}
\dot X=AX+B_xXu_x+B_yXu_y+B_zXu_z,
\ee
with $$A:=-i
\sum_{k<l, k,l=1}^n
(M_{kl} I_{kx,lx}+ N_{kl} I_{ky,ly}+ P_{kl} I_{kz,lz}),
$$ and
$$B_{v}:=-i (\sum_{k=1}^n
\gamma_k I_{kv}), \ \ \text{ with } v=x,\, y, \text{ or } z.$$
 It is clear that the
controllability properties of this
class of systems only depends on the
parameters ${M_{kl},N_{kl},P_{kl}}$ and $\gamma_k$.
Our goal in the next sections is
to characterize the structure of the Lie algebra generated by $A$ and $B_x$, $B_y$, $B_z$, 
$\cal L$,  in terms of
these parameters. The network of spin particles  can be represented by a
graph whose nodes represent the particles and are labeled by their
gyromagnetic ratios and an edge connects the nodes corresponding to
particles $k$ and $l$ if and only if at least one of the coupling 
constants ${M_{kl},N_{kl},P_{kl}}$ is different from zero. In this case, the edge
is labeled by the triple $\{ M_{kl}, N_{kl}, P_{kl} \}$. 
 It is our goal,
in the next sections, to relate the properties of the Lie algebra $\cal
L$, generated by $A$, $B_x$, $B_y$ and $B_z$ to the properties of this graph. 
In the following, we denote this graph by $\cal Gr$.

\vs

We define an ordering on the $n$ particles so that the first $n_1$
have the same gyromagnetic ratio $\gamma_1$, the next $n_2$
particles all have gyromagnetic ratio $\gamma_2$, with $\gamma_2
\not= \gamma_1$, and so on up to the $r-$th set of $n_r$ particles
with gyromagnetic ratio $\gamma_r$, with $\gamma_j \not= \gamma_k$ when 
$j \not= k$ and
$n_1+n_2+n_3+ \cdot \cdot \cdot + n_r=n$. We shall denote  the first
set of particles by $S_1^0$, the second one by $S_2^0$, and so on up to
the $r-$th, $S_r^0$. 
We also define, for $j=1,2,...,r$,  $v=x,y,z$
\be{ti}
\tilde{I}_{jv}: = \sum_{h\in S^0_j} I_{hv}, 
\ee
and we have 
$$
B_v := -i \sum_{j=1}^r\gamma_j \tilde{I}_{jv}.
$$
For a given system, we shall call the {\it Control Subalgebra} of $\cal
L$, the subalgebra generated by the matrices $B_x$, $B_y$ and $B_z$. We
shall denote the control subalgebra by $\cal B$. 
%Using the first of the  previous equalities and the fact that the
%sets $S^0_j$ are disjoint it is not difficult to 
%show that, 
%for the matrices $\tilde{I}_{k_1v_1,\ldots,k_lv_l}$, both
%Properties 1 and 2, given for $I_{k_{1}v_{1},\ldots, k_{l}v_{l}}$
%in Appendix $A$, hold.

\section{Characterization of the Control Subalgebra}
\label{CCS}

The following lemma shows that the control subalgebra ${\cal B}$ of a
spin system is the direct sum of $r$ subalgebras isomorphic to $su(2)$. 
\vs

\bl{6l} Assume we are given a model as in (\ref{scrod}), and let
$\gamma_1,\ldots,\gamma_r$ be the different values for the
gyromagnetic ratios. Assume that to each value $\gamma_j$
correspond $n_j$ particles in the set $S^{0}_{j}$, $j=1,\ldots,r$, 
then the matrices $B_x$, $B_y$ and
$B_z$ generate the following Lie algebra:
\be{BBBtilde}
{{\cal B}}={ {\cal B}}_x \oplus { {\cal B}}_y
\oplus  { {\cal B}}_z, \ee with: \be{Bxtilde}
{{\cal{B}}}_x\, =\, span_{j=1,\ldots,r}\{i \tilde{I}_{jx}\},
\ee
 \be{Bytilde}
{{\cal{B}}}_y\, =\, span_{j=1,\ldots,r}\{i \tilde{I}_{jy}\},
\ee
 \be{Bztilde}
{{\cal{B}}}_z\, =\, span_{j=1,\ldots,r}\{i \tilde{I}_{jz}\}.
\ee Moreover, we have:
 \be{commutazzz} [{ {\cal B}}_x, { {\cal
B}}_y] = { {\cal B}}_z,  \qquad [{ {\cal B}}_y, {
{\cal B}}_z] ={  {\cal B}}_x,  \qquad [{ {\cal B}}_z,
{ {\cal B}}_x] = { {\cal B}}_y. \ee \el 
\bpr First, notice that $\tilde I_{j(x,y,z)}$ satisfy the commutation
relations 
\be{commutilde}
[\tilde I_{jx}, \tilde I_{ky}]=i \delta_{jk} \tilde I_{jz}, \quad 
[\tilde I_{jy}, \tilde I_{kz}]=i \delta_{jk} \tilde I_{jx}, \quad 
[\tilde I_{jz}, \tilde I_{kx}]=i \delta_{jk} \tilde I_{jy},   
\ee
where we used the Kronecker symbol $\delta_{jk}$. We
proceed by induction on $r\geq 1$. If $r=1$, then we have, for
$v\in \{x,\ y\ z\}$:
\[
B_v\, =\, -i \tilde{I}_{1v},
\]
thus (\ref{BBBtilde})-(\ref{commutazzz}) follow immediately {}from
the basic commutation relations (\ref{commutilde}).

To prove the inductive step, we first  show, again by induction on
$r\geq 1$ that:
 \be{risultat}
 \begin{array}{lcl}
 [B_x,B_y] & = & -i \sum_{j=1}^r \gamma_j^2 \tilde{I}_{jz},
  \\
{[}B_y,B_z{]} & =  & -i \sum_{j=1}^r \gamma_j^2 \tilde{I}_{jx},
 \\
 {[}B_z,B_x] & = & -i \sum_{j=1}^r \gamma_j^2 \tilde{I}_{jy}.
\end{array}
 \ee
We will prove only the first of the previous equalities, since 
the
other ones may be obtained in the same way. If $r=1$, then
\[
[B_x,B_y]=- \gamma_1^2[ \tilde{I}_{1x},\tilde{I}_{1y}]=
-i\gamma_1^2 \tilde{I}_{1z},
\]
where to get the last equality we have used (\ref{commutilde}).
Now let $r>1$: \[
[B_x,B_y]= -[ \sum_{j=1}^{r}\gamma_j\tilde{I}_{jx},
\sum_{j=1}^{r}\gamma_j\tilde{I}_{jy} ] = \] \[- \left( [
\sum_{j=1}^{r-1}\gamma_j\tilde{I}_{jx},
\sum_{j=1}^{r-1}\gamma_j\tilde{I}_{jy} ] + \sum_{j=1}^{r-1} [
\gamma_j\tilde{I}_{jx}, \gamma_r\tilde{I}_{ry}] + \sum_{j=1}^{r-1}
[ \gamma_r\tilde{I}_{rx}, \gamma_j\tilde{I}_{jy}] + [
\gamma_r\tilde{I}_{rx}, \gamma_{r}\tilde{I}_{ry}] \right).
\]
By the inductive assumption, we have: \be{uno}
[\sum_{j=1}^{r-1}\gamma_j\tilde{I}_{jx},
\sum_{j=1}^{r-1}\gamma_j\tilde{I}_{jy} ]= i \sum_{j=1}^{r-1}
\gamma_j^2 \tilde{I}_{jz}. \ee Using (\ref{commutilde}),  we obtain, 
for $j<r$,  \be{due}
\begin{array}{l}
[\gamma_j\tilde{I}_{jx}, \gamma_r\tilde{I}_{ry}]=0,
\\
{[} \gamma_r\tilde{I}_{rx}, \gamma_j\tilde{I}_{jy}]=0, 
\end{array}
\ee 
and 
\be{tre} [ \gamma_r\tilde{I}_{rx}, \gamma_{r}\tilde{I}_{ry}]= i
\gamma_r^2 \tilde{I}_{rz}. 
\ee 
Now putting together equations
(\ref{uno}), (\ref{due}) and (\ref{tre}), we get:
\[
[B_x,B_y]\, =\,- i \sum_{j=1}^{r} \gamma_j^2 \tilde{I}_{jz},
\]
as desired. Thus, we have proved (\ref{risultat}).

Now notice  that, for example,  $[B_y,B_z]$ has the same form as
$B_x$ except that the $\gamma_j$'s have been replaced by
$\gamma_j^2$, therefore, using the same arguments as above one may
show that:
 \be{calculatio} [[B_y,B_z], B_y]=-i
\sum_{j=1}^r \gamma_j^3 \tilde I_{jz}. \ee More in general,
considering the Lie bracket between $F_{x}:=-i\sum_{j=1}^r
\gamma_j^k \tilde I_{jx}$, and $G_{y}:=-i\sum_{j=1}^r \gamma_j^l
\tilde I_{jy}$, we get $S:=-i\sum_{j=1}^r \gamma_j^{k+l} \tilde
I_{jz}$. Proceeding this way,  we obtain all the
matrices \be{matre} i \sum_{j=1}^r \gamma_j^l \tilde I_{jx}, \ee
\be{matrey} i \sum_{j=1}^r \gamma_j^l \tilde I_{jy}, \ee and
\be{matrez} i \sum_{j=1}^r \gamma_j^l \tilde I_{jz}, \ee
$l=1,...,r$. The matrices in (\ref{matre}) form a basis in $
{\cal B}_x$ since the $\tilde I_{jx}$ do and the linear
transformation in (\ref{matre}) is nonsingular. In fact, the
corresponding determinant is a Vandermonde determinant which is
different {}from zero because all the $\gamma_j$'s are different
{}from each other. The same is true for the elements in
(\ref{matrey}) and  (\ref{matrez}) which form a basis in 
${\cal B}_y$ and $ {\cal B}_z$, respectively. Finally, the
commutation relations (\ref{commutazzz}) follow  immediately {}from
 (\ref{commutilde}). \epr

\vs

Notice that it follows {}from (\ref{commutilde}) and (\ref{commu}) 
that the subalgebras spanned by $\tilde I_{j(x,y,z)}$ are each isomorphic
to $su(2)$ and they commute with each other. 
For a given $j$, the Lie group corresponding to 
$span \{ I_{j(x,y,z)} \}$ is given by $n_j$
copies of $SU(2)$   (where $n_j$ denotes denotes 
the number of particles with gyromagnetic ratio
$\gamma_j$) \footnote{This is the Lie group of matrices of the
form $I_1 \otimes L \otimes I_2$, where the 
identity matrix $I_1$ has dimension $2^{n_1+\cdot \cdot \cdot n_{j-1}}$,
the identity matrix $I_2$ has dimension 
$2^{n-n_1-n_2-\cdot \cdot \cdot n_j}$ and $L$ has dimension $2^{n_j}$ 
and is  has the form 
$F \otimes F \otimes \cdot \cdot \cdot \otimes F$, with $F \in SU(2)$ and the Kronecker product
having $n_j$ factors.}. Therefore it is isomorphic to $SO(3)$ or $SU(2)$ according to whether $n_j$ is even or
odd, respectively.     

\section{Lie Algebra Structure and Controllability with Different
Gyromagnetic Ratios}  
\label{las}
%%%%%%%%%%%%%%%%%%%%%%%%%%%%%%%%%%%%%%%%%%%%%%%%%%%%%%
In this section, we shall assume
that the gyromagnetic ratios
$\gamma_1,...,\gamma_n$ are all
different. Therefore we have $r=n$
and, {}from Lemma \ref{6l}, we have that the control subalgebra $\cal
B$ is the span of the $iI_{j(x,y,z)}$, $j=1,...,n$. We shall 
give a necessary and sufficient condition of controllability and
describe the nature of the Lie algebra $\cal L$, in terms
of the properties of the graph $\cal Gr$. This 
graph will, in
general,  have a number $s$ of 
connected components. We first
describe the situation when $s=1$ 
and then generalize to the case
of arbitrary $s$.

\vs

\bt{7l} Assume we are given a
model as in (\ref{scrod}), where the
values  $\gamma_j$,
$j=1,\ldots,n$
of the  gyromagnetic ratios are all different.
 If the graph
$\cal {G} \cal r$ is connected, then
\be{dbk} {\cal
L}=su(2^{n}).  
\ee
As a consequence the system 
is operator and state controllable (see Theorem \ref{sommarize}). 
\et
\bpr We show that all the matrices of the
form $iI_{k_1 l_1, k_2 l_2,...,k_m l_m}$ can be obtained as
repeated commutators of $A$, $B_x$, $B_y$, $B_z$, for every $1
\leq m  \leq n$. Lemma \ref{6l} gives the result for $m=1$. We
first prove that this is true for $m=2$ as well, and then proceed
by induction on $m$. If $m=2$, we want to show that we can obtain
all the matrices of the form $iI_{kv,lw}$, $k <l$, $v,\ w \in
\{x,\ y,\ z\}$. {}From our assumption on the connectedness of
$\cal G \cal r$, there exists a path joining the node representing the
$k-th$ particle and the node representing the $l-$th particle. Let
us denote by $p$ the {\it length} of this path, namely the
 number of edges between $k$ and $l$. We proceed by induction on $p$.
If $p=1$, then at least one among $M_{kl},$ $N_{kl}$ 
and $P_{kl}$ is different from zero. If $P_{kl} \not=0$, 
we have:  \be{fty} [A,iI_{lx}]= i \left(
\sum_{h<l}(-N_{hl}I_{hy,lz} + P_{hl}I_{hz,ly}) +
\sum_{h>l}(-N_{lh}I_{lz,hy} + P_{lh}I_{ly,hz}), \right) \ee and \be{gh}
[[A,iI_{lx}], -iI_{ky}]=-iP_{kl} I_{kxly}. \ee Since $P_{kl}
\not=0$, {}{}from the matrix $-iP_{kl} I_{kxly}$, using (repeated)
Lie brackets with elements $iI_{kf}$ and/or $iI_{lf'}$, with $f,\
f'\in \{x,\ y,\ z\}$ one can obtain all of the elements of the
form $iI_{kv,lw}$, with $v, \ w \in \{x,\ y,\ z\}$. If $P_{kl}=0$, but 
$N_{kl} \not =0$, the same can be proved by taking the commutator with 
$iI_{lx}$ first and then the commutator with $iI_{kz}$ and analogously, 
if $N_{kl}=P_{kl}=0$, by taking the commutator with $iI_{ly}$ first and
then with $iI_{kz}$.  Now, assume it
is possible to obtain every $iI_{kv,lw}$ for every $k<l$ whose
distance is $\leq p-1$. Let $k$ and $l$ have a path with distance
$p$ and let $\bar l$ represent a particle/node in between $k$ and
$l$ in the path. Let us also assume just for notational
convenience that $k<\bar l <l$. {}From the inductive assumption,
we know that $iI_{kv, \bar l w}$ and $iI_{\bar l f, lf'}$ can be
obtained for every $v,\, w,\, f,\, f' \in \{x,\ y,\ z\}$. We need
to show that we can also obtain every $iI_{kg,lq}$ for every $g,\,
q\in \{x,\ y,\ z\}$. Using equation (\ref{Prop2bis}) in Appendix $A$, 
we get
\be{altroconto} [iI_{kx,\bar l x}, -i I_{\bar l y,
ly}]=iI_{kx,\bar l z, ly}, \ee and \be{ac1} [iI_{kx,\bar l z, ly},
i I_{\bar l z, lx}]=\frac{1}{4} i I_{kx,lz}, \ee where we have
used the following property of the Pauli matrices \be{PP}
\sigma_{x}^2=\sigma_y^2=\sigma_z^2=\frac{1}{4} I_{2 \times 2}. \ee
As before, we can now take repeated Lie brackets of the matrix
obtained in (\ref{ac1}) with matrices of the form $iI_{kf}$ and/or
$iI_{lf'}$, with $f,\, f' \in \{x,\ y,\ z\}$, to obtain all of the
matrices $iI_{kv,lw}$, for $v,\, w\in \{x,\, y,\, z\}$. This
concludes the proof that every Kronecker product with two matrices
different {}from the identity can be obtained, namely $m=2$ in the
above notations.

We now show that every matrix
$iI_{k_1v_1,k_2 v_2,...,k_mv_m}$ can
be obtained. Consider the Lie bracket
\be{anotherLie} [-iI_{k_1
v_1, k_2 v_2,...,k_{m-1} x}, i I_{k_{m-1} y, k_m v_m}]= iI_{k_1
v_1, k_2 v_2,...,k_{m-1} z, k_m v_m}.
\ee Both elements
$-iI_{k_1
v_1, k_2 v_2,...,k_{m-1} x}$ and $i I_{k_{m-1} y, k_m v_m}$ are
available because of the inductive assumption. If $v_{m-1}=z$, we
have concluded otherwise,  the Lie bracket with the
matrix $iI_{k_{m-1}x}$ or $iI_{k_{m-1}y}$ leads to the desired
result. This conclude the proof of the Theorem. \epr

In the general situation,
 assume that $\cal G \cal r$ has $s$ connected
components and denote by $n_j$ the number of nodes in the $j-$th
component. Set up an ordering of the particles so that the first $n_1$
are in the first connected component of the graph,
 the ones {}from $n_1+1$
up to $n_1 +n_2$ are in the second component and so on. We
have $n_1+n_2+\cdot \cdot \cdot n_s=n$. The following
Theorem describes
the structure of the Lie algebra $\cal L$ in the general case assuming
to have different gyromagnetic ratios $\gamma_i$, $i=1,2,...,n$.
\bt{8} Assume we are given a model
as in (\ref{scrod}), where the
values  $\gamma_j$, $j=1,\ldots,n$,
of the  gyromagnetic ratios
are all different. Moreover,  assume 
that the graph $\cal G \cal r$ has
$s$ connected components (as described above), then
 \be{elle} {\cal
L}= {\cal S}_1 \oplus {\cal S}_2 \oplus \cdot \cdot \cdot \oplus
{\cal S}_s, \ee where each ${\cal S}_j$, $j=1,2,...,s$ is the
subalgebra spanned by the matrices \be{sdfh} iI_{k_1v_1,k_2
v_2,...,k_r v_r}, \ee with \be{with} n_1+n_2+ \cdot \cdot \cdot
+n_{j-1}<k_1< k_2 < \cdot \cdot \cdot < k_r \leq n_1+n_2+ \cdot
\cdot \cdot +n_{j}. \ee \et \bpr First notice that, {}from equation
(\ref{Prop1}) in Appendix A, it follows immediately:
 \be{commul} [{\cal S}_ j,
{\cal S}_k]=0, \ \ \text{ if } j\neq k. \ee Since the values
$\gamma_j$ are all different, {}from Lemma \ref{6l} we have that all
the elements
 of the form $iI_{kv}$, $k=1,\ldots,n$, $v\in\{x,\, y,\, z\}$, are in
$\cal L$. We can write  the matrix $A$ as
\be{mtra}
\begin{array}{lll}
A&=&-i(\sum_{1\leq k<l \leq n_1}  (M_{kl} I_{kxlx} +
N_{kl}I_{kyly}+
P_{kl}I_{kzlz})+ \\
&
+
&
\sum_{n_1 <k<l \leq n_1+n_2}  (M_{kl}I_{kxlx}
+N_{kl}I_{kyly}+ P_{kl}I_{kzlz})+\\
 &
\cdot \cdot \cdot &
+
\sum_{n_1+n_2+ \cdot \cdot \cdot n_{s-1} <k<l \leq n} 
(M_{kl}I_{kxlx} +N_{kl}I_{kyly}+P_{kl} I_{kzlz}),
\end{array}
\ee
using the fact that $M_{kl}=N_{kl}=P_{kl}=0$ 
if $k$ and $l$ are in two different
connected components. Taking  the
Lie brackets with elements $iI_{kv}$, $v\in \{
x,\, y,\, z\}$, with $n_1+n_2 + \cdot \cdot \cdot n_{j-1}<k\leq
n_1+n_2 + \cdot \cdot \cdot n_{j}$ (here if $j=1$, we put
$n_0=0$), one may show, as in 
the proof of Theorem \ref{7l}, that it is possible to obtain  all
the elements in ${\cal S}_j$, $j=1,2,...,s$.
Moreover {}from (\ref{commul}), it follows that these and their
linear combinations are the only matrices that can be generated by
$A$, $B_x$, $B_y$, $B_z$. \epr

Notice that, in the above situation, one may think of the spin
system as a parallel connection of $s$ spin systems of dimension
$n_j$, $j=1\ldots,s$, controlled in parallel by the same control.
The solution of (\ref{scrod}) has the form
\be{formasol}
X(t)=\Phi_1(t)\Phi_2(t)\cdot \cdot \cdot \Phi_s(t),
\ee
where
$\Phi_j(t)$ is the solution of (\ref{scrod}) with
\be{newA}
A=-i\sum_{n_{j-1}<h<k\leq n_j}  (M_{hk}I_{hx,kx}+ 
N_{hk} I_{hy,ky}+ P_{hk} 
I_{hz,kz}), \ee and \be{bxbybz} 
B_{v}=
-i \sum_{k=n_{j-1}+1}^{n_j}
\gamma_k I_{kv}, \ \ \ v\in \{x,\, y,\, z\}.
\ee The controls are the same for every subsystem and the matrices
$\Phi_j$ in (\ref{formasol}) 
commute due to (\ref{commul}). The set of states that can be obtained 
with an appropriate
control for system (\ref{scrod}) is given by the Lie
group corresponding to the Lie 
algebra $\cal L$ namely, in this case,  
$SU(2^{n_1}) \otimes 
SU(2^{n_2}) \otimes \cdot \cdot \cdot \otimes SU(2^{n_s})$. 

\br{nuovo}
It is important to notice, and it will be used later 
in the next Section, that, in 
Theorems \ref{7l} and \ref{8}, the assumption of different 
gyromagnetic ratios   is used only 
to derive  that the Lie algebra
spanned by $iI_{j(x,y,z)}$ is a subalgebra of $\cal L$. Thus both 
statements of Theorems \ref{7l} and 
\ref{8} remain true if, instead 
of assuming $\gamma_{i}\neq \gamma_{j}$ for all $i\neq j$, we assume 
$span_{j=1,...,n} \{ iI_{j(x,y,z)} \}  \subseteq {\cal L}$. This fact
will be used in the following Section. 
\er

\vs

In the following Theorem, we answer  the question of 
state controllability for 
spin systems with different gyromagnetic
ratio. It follows from Theorem  \ref{sommarize} that,  
if ${\cal L}=su(2^n)$, the set of states reachable
for system (\ref{scrod}) is $SU(2^n)$ and therefore the system is both
operator controllable and state controllable in this case. If 
${\cal L} \not=su(2^n)$,we have seen that the set of states reachable
for (\ref{scrod})  is $SU(2^{n_1}) \otimes 
SU(2^{n_2}) \otimes \cdot \cdot \cdot \otimes SU(2^{n_s})$. To see that the
system is not state controllable, notice that the corresponding Lie
algebra $\cal L$ is not {\it simple} 
(since each of the subalgebras isomorphic to $su(2^{n_j})$ is actually 
an ideal in $\cal L$) and therefore it cannot be isomorphic to
$sp(2^{n-1})$ as in Theorem 1, part (3). A more direct and geometric
proof of the fact  that  $SU(2^{n_1}) \otimes 
SU(2^{n_2}) \otimes \cdot \cdot \cdot SU(2^{n_s})$ is not transitive on
the complex sphere is to reason as follows. Assume for
simplicity $s=2$ and $V_1$ and $V_2$ two subspaces,  
of dimension  $2^{n_1}$ and $2^{n_2}$ such that the underlying subspace
of the overall system is $V_1 \otimes V_2$. 
Every `unentangled' state, namely a
state of the form $|v_1> \otimes |v_2>$,  
with vectors $|v_1> \in V_1$
and $|v_2> \in V_2$ can only be 
transformed into another unentangled vector 
$(A \otimes B) (|v_1> \otimes |v_2>)= A | v_1> \otimes B | v_2>$ and
there is no possibility of transforming $|v_1> \otimes |v_2>$ into an
entangled vector namely a vector that cannot be written as the tensor
product of two vectors {}from $V_1$ and $V_2$. On the other hand,
entangled states  always exist for a pair of non trivial vector spaces
$V_1$ and $V_2$ (for example, if $|e_j>$, $j=1,...,m_1$, is a basis of 
$V_1$ and $|f_k>$, $k=1,...,m_2$ is a basis of $V_2$, so that 
$|e_j>|f_k>$ is a basis of $V_1 \otimes V_2$, consider 
$\frac{1}{\sqrt{2}} |e_1>|f_1>+\frac{1}{\sqrt{2}}|e_{m_1}>|f_{m_2}>$.) We
summarize the results in this section with the following theorem.

\bt{9} Consider a system of $n$-spins with different gyromagnetic
ratios given by the model (\ref{scrod}). For this system all the
controllability notions are equivalent and they are verified if
and only if the associated graph $\cal G \cal r$ is connected. \et

\br{RDM}

In many physical implementations of the control of spin $\frac{1}{2}$ 
particles,  the $z$ component of the control is held constant. The 
only changes in the previous treatment occur in the proof of Lemma 
\ref{6l}. In fact, for this case, one does not have the matrix $B_{z}$. 
However, by using 
the first one of equations 
(\ref{risultat}), one obtains $-i \sum_{j=1}^{r}\gamma_{j}^{2}\tilde 
I_{jz} \in  {\cal B}$. Then, using this matrix in place of 
$B_{z}$, one gets all the matrices in (\ref{matre}), (\ref{matrey}), 
(\ref{matrez}), with only odd $l$'s in (\ref{matre}), (\ref{matrey}), 
and even $l$'s in (\ref{matrez}). If we assume 
$|\gamma_{1}| \not= |\gamma_2| \not= \ldots|\gamma_{r}|$, 
the result remains unchanged. 
In fact, the determinant of the matrix referred to at the end of the 
proof of Lemma \ref{6l}, is still a non zero Vandermonde determinant.
The 
drift matrix $A$ is modified by adding a term 
$-i\sum_{j=1}^{n}\gamma_{j}I_{jz}u_{z}$, with $u_{z}$ constant but 
this does not modify the resulting Lie algebra $\cal L$, since 
$-i\sum_{j=1}^{n}\gamma_{j}I_{jz}u_{z}$ belongs to the control 
subalgebra.
\er

\section{Systems with Possibly Equal  Gyromagnetic Ratios}

\label{SPEGR}
In this section we analyze the graph $\cal Gr$ for networks of spins with
possibly equal gyromagnetic ratios and give a sufficient condition of
 operator controllability for these systems in terms of the properties
of this graph. It will follow from the examples in the next section that
the equivalence between state controllability and operator
controllability, proved in Theorem \ref{9} for systems with different
gyromagnetic ratios,  does not always hold if we allow two particles to
have the same gyromagnetic ratio. 

%As pointed out in Remark \ref{nuovo} all we have to prove to obtain
%operator controllability is that the Lie algebra $span
%\{iI_{j(x,y,z)} \}$ is a subalgebra of $\cal L$. We do this in the
%following Theorem. 
In the following we describe an algorithm on the graph $\cal Gr$ to
conclude operator controllability. 
The main idea and the physical interpretation go as
follows.  When all the gyromagnetic
ratios of the particles are different they `react' in a
different way to the common electro-magnetic field and this
`asymmetry' along with connectedness of the spin network 
allows us to control all the particles at the same time.
However, even if two particles have equal gyromagnetic ratios they
might interact in different ways with a third particle which has
gyromagnetic ratio different from the two, and this will break
once again the symmetry and give controllability.

\vs

Let us  divide
the particles into $r$ sets $S_1^0,...,S_r^0$ as it was done in
Section 3 and assume that at least one    set is a  singleton,
namely, there exists at least one particle which has different
$\gamma$ {}from all the others.  Consider a set $\bf S$ containing
all the singleton nodes. Assuming that there are $m$ of them, let the
sets $S^0_{1},$...,$S^0_{r-m}$ be of cardinality $\geq 2$. 
Now we illustrate a `disintegration'
procedure to divide these sets further.

\vs

\noindent{\bf Algorithm 1} 

\begin{enumerate}

\item Let ${\cal C}$ be a collection 
of sets. Set ${\cal C}:=S^0_{1}, S^0_{2},...,
S^0_{r-m}$.  

\item For each  set $\tilde S$ in $\cal C$, consider a particle 
$\bar l$ in $\bf S$ such that 
 for at least two particles in $k$ and $j$ in $ \tilde S$
\be{conditionA}
\{|M_{k \bar l}|,|N_{k \bar l}|, |P_{k \bar l}| \} \not= 
\{|M_{j \bar l}|,|N_{j \bar l}|, |P_{j \bar l}| \}. 
\ee 
If there is no
element in $\bf S$ and no set in $\cal C$ having this property STOP. 
Divide the set $\tilde S$ into subsets of particles that have the
same value for $\{|M_{k \bar l}|,|N_{k \bar l}|, |P_{k \bar l}| \}$.

\item Consider the sets obtained in Step  2. Put the 
elements that are in
singleton sets in $\bf S$. If all the elements are in $\bf S$, STOP.

\item Replace the collection $\cal C$ with the remaining non singleton
sets and go back to Step 2.  

\end{enumerate}

\vs

We have the following theorem.

\bt{10}
If Algorithm 1 ends with all the particles in the set
$\bf S$ and $\cal Gr$ is connected, then 
the Lie algebra $\cal L$ associated to the spin
$\frac{1}{2}$ particles system, with $n$ particles, 
is $su(2^n)$. 
As a consequence the system is operator controllable. 
More in general, if Algorithm 1 ends with all the 
particles in the set
$\bf S$ and $\cal Gr$ has $s$ connected 
components of cardinality $n_1$, $n_2$, ..., $n_s$, $\cal L$
is given by (\ref{elle})-(\ref{with}) (See Theorem \ref{8}).  \et
\bpr
{}From Remark  \ref{nuovo}, all we have to show is
that, in the given situation, the Lie algebra $span_{j=1,...,n}
\{iI_{j(x,y,z)} \}$ is a
subalgebra of $\cal L$. 
Rewrite the drift matrix $A$ as
\begin{eqnarray}
A=-i\sum_{k<l, k \notin S_{r-m}^0, l \notin S_{r-m}^0} 
(M_{kl}I_{kx,lx}+N_{kl}I_{ky,ly}+P_{kl}I_{kz,lz}) 
\nonumber \\ -i \sum_{k<l, k \in
S_{r-m}^0, l \in S_{r-m}^0}
(M_{kl}I_{kx,lx}+N_{kl} I_{ky,ly}+
 P_{kl}I_{kz,lz})
\nonumber \\ -i \sum_{k<l, k \in S_{r-m}^0, l \notin S_{r-m}^0}
 (M_{kl}I_{kx,lx}+ N_{kl} I_{ky,ly}+
 P_{kl} I_{kz,lz})  \nonumber \\ -i \sum_{k<l,
k \notin S_{r-m}^0, l \in S_{r-m}^0} 
(M_{kl} I_{kx,lx}+ N_{kl} I_{ky,ly}+ P_{kl}I_{kz,lz}). \label{formA}
\end{eqnarray}
{}From Lemma \ref{6l}, the matrices $i\tilde I_{jv}$, $v\in \{x,\,
y\, z\}$ and  $j=1,2,...,r$, where $r$ is the number of sets
$S^0_j$, are available to generate the Lie algebra $\cal L$. In
particular, since we have assumed that the last $m$ 
sets are singletons, the matrices $iI_{lv}$, $v \in \{x,y,z\}$,
$l=n_1+n_2+ \cdot \cdot \cdot + n_{r-m}+,...,n$ 
are $\in {\cal L}$. Now, assume that in the set 
$S_{r-m}^0$ there are two elements $j$ and $k$ such that condition
(\ref{conditionA}) is verified for some $  \bar  l \in {\bf S}$ and assume, for
the sake of concreteness, that the inequality is verified for the $P$
coefficient (minor changes are needed in the other cases). By
taking the Lie bracket of $A$ with $i\tilde I_{(r-m)x}$, the first
term gives zero, since it does not involve any term in the set
$S^0_{r-m}$ (see (\ref{Prop1}), in Appendix $A$ and the definition of
the $\tilde I$'s in (\ref{ti})). The Lie bracket of the second term with 
$i\tilde I_{(r-m)x}$ gives a matrix which is a linear combination 
of matrices of the form $iI_{kv,pw}$, $k,p \in S^0_{r-m}$ and $v,w \in
\{x,y,z\}$. We call this matrix $K_{r-m}$.  
Thus, we have \be{clopl}\begin{array}{l}
[A,i\tilde I_{(r-m)x}]=K_{r-m}+ 
\\ i \left( \sum_{k<l, k \in S_{r-m}^0, l
\notin S_{r-m}^0}  (-N_{kl}I_{kz,ly}+P_{kl}I_{ky,lz}) 
+ \sum_{k<l, k
\notin S_{r-m}^0, l \in S_{r-m}^0} 
(-N_{kl} I_{ky,lz}+ P_{kl}I_{kz,ly})\right). \end{array}  \ee 
%Now,
%consider a particle $k$ such that  
%there exists a particle $\bar l$ in one of the  singleton sets 
%$S^0_{r-m+1},\ldots,S^0_r$, namely in $\bf S$, such that 
%\be{diseg} |P_{k \bar l}|
%\not= |P_{j \bar l}|, \ee for every pair of particle $j$ 
%in $S^0_{r-m}$. 
By taking the Lie bracket of (\ref{clopl}) with
$iI_{\bar ly}$, and  using Properties 
1 and 2 in the Appendix $A$,
we obtain 
\be{what} 
[[A,i \tilde I_{(r-m)x}],iI_{\bar l y}]= 
i \sum_{k \in S_{r-m}^0} P_{k \bar l} I_{k y,
\bar lx}. \ee {}From this matrix, by taking Lie brackets with $i
\tilde I_{(r-m)v}$ and/or $iI_{\bar lv}$, $v\in \{x,\, y\, z\}$,
it is possible to obtain all the matrices of the form (\ref{what})
with all the possible combinations of $x,y$ and $z$ in place of
$y$ and $x$ respectively.

Using (\ref{fundapro}) in Appendix A, 
it is not difficult to see that
\be{LB}
[i \sum_{k \in S_{r-m}^0} P_{k \bar l}i I_{k y, \bar l z}, i
\sum_{k \in S_{r-m}^0} P_{k \bar l}i I_{k x, \bar l z}]=
\frac{1}{4}i\sum_{k \in S^0_{r-m}} P_{k \bar l}^2 I_{kz}. \ee By
taking the Lie bracket of this with $-i \sum_{k \in S_{r-m}^0}
P_{k \bar l}i I_{k x, \bar l z}$, we obtain $i \sum_{k \in
S_{r-m}^0} P_{k \bar l}^3i I_{k y, \bar l z}$ and repeating the
calculation as in (\ref{LB}), we obtain \be{LBplus} [i \sum_{k \in
S_{r-m}^0} P_{k \bar l}^3i I_{k y, \bar l z}, i \sum_{k \in
S_{r-m}^0} P_{k \bar l}i I_{k x, \bar l z}]= \frac{1}{4} i\sum_{k
\in S^0_{r-m}} P_{k \bar l}^4 I_{kz}. \ee Continuing this way, it
is possible to obtain all the matrices of the form \be{fom}
i\sum_{k \in S^0_{r-m}} P_{k \bar l}^{2p} I_{kz}, \qquad
p=0,1,2,..., \ee and, with minor changes in the choice of the 
Lie brackets, we can obtain
\be{foman} i\sum_{k \in S^0_{r-m}} P_{k \bar l}^{2p} I_{kx}, \ \
i\sum_{k \in S^0_{r-m}} P_{k \bar l}^{2p} I_{ky}, \qquad
p=0,1,2,.... \ee 
Now consider for example, the matrices 
$i\sum_{k \in S^0_{r-m}} P_{k \bar l}^{2p} I_{kz}$ and assume, without
loss of generality that the elements $P_{k\bar l}^{2p} I_{kz}$ are
arranged so that  elements that have the same value for $P_{kl}$ appear one
after the other in the sum. The associated determinant is (cfr. the
proof of Lemma \ref{6l}) a Vandermonde determinant and therefore by
appropriate linear combinations we can obtain all the matrices of the
form $\sum_{k \in T} I_{kz}$ where $T$ is a generic subset of
$S^0_{r-m}$ such that all the values of $|P_{k \bar l}|$ are the same, 
for all the $k \in T$. In particular, if $T$ contains a single element
then  we place that element in the set of singletons $\bf S$. The other
subsets of $S^0_{r-m}$ are arranged in  new sets. It is clear that we
can repeat this procedure for the other sets $S^0_1,
S^0_2,...,S^0_{r-m-1}$, and then for the subsets obtained, as described
in Algorithm 1.  If the procedure ends with all the elements in $\bf S$
then we have that 
$span_{j=1,...,n} \{ iI_{j(x,y,z)} \}$ is in $\cal L$ and the Theorem follows 
from Remark
(\ref{nuovo}). 
\epr

\section{Low dimensional systems}
\label{lds}

Results on the controllability  of spin systems in the 
cases of  $n=1$ and 
$n=2$ particles can be found in 
\cite{ioSCL}, \cite{HOMO} and \cite{BKS}.  In this section we consider 
 the model (\ref{scrod}) assuming Heisenberg type of interaction namely 
\be{isotropic}
M_{kl}=N_{kl}=P_{kl}:= J_{kl}, 
\ee
for every pair of particles $k$ and $l$. For this model, in the 
case $n=2$, the only
noncontrollable case, ${\cal L} \not= su(2)$, is when 
$n=2$ and the two particles have the same gyromagnetic ratio. In this 
situation, we have 
\be{lio}
{\cal L}=span\{A\} \oplus span\left\{i\left(\sigma_{v}\otimes I_{2\times 2}+
I_{2\times 2} \otimes \sigma_{v}\right), \quad v\in \{x,y,z\}\right\}, 
\ee
and the matrix $A$ commutes with all the matrices in $\cal L$. The 
 Lie algebra $\cal L$ is isomorphic to $u(2)$. 

\vs 

We treat now completely the case of $n=3$ interacting spin 
$\frac{1}{2}$ particles. If the three particles have all different 
gyromagnetic ratios, then we are in the situation treated in Section
\ref{las}.  
There are  two more 
possibilities:
\bi
\item[(a)]
all the three gyromagnetic ratios all equal (i.e. 
$\gamma_{1}=\gamma_{2}=\gamma_{3}$),
\item[(b)]
two gyromagnetic ratios are equal 
and the third is different (i.e. 
$\gamma_{1}=\gamma_{2}$ and 
$\gamma_{1}\neq\gamma_{3}$, according to 
the notations in Section \ref{misp}, we have 
 $S^{0}_{1}=\{ 1,2 \}$ and $S^{0}_{2}=\{3\}$). 
\ei

\vs
{$\bullet$ {\bf{case (a)}}}

This case is particularly simple. In fact, we have:
\be{lio1}
{\cal L}=span\{A\} \oplus span\{i\tilde{I}_{1v}, \quad v\in \{x,y,z\}\}, 
\ee
with 
\[
\left[ span\{A\}, span\{i\tilde{I}_{1x},i\tilde{I}_{1y},i\tilde{I}_{1z}\}
\right]=0.\]
The Lie algebra $\cal L$ is isomorphic to $u(2)$ and 
the model is neither operator controllable nor state controllable from
Theorem \ref{sommarize}.

\vs
{$\bullet$ {\bf{case (b)}}}

This situation is more involved and it gives rise to interesting 
examples. First recall that, {}from  Lemma \ref{6l}, we get, for 
$v=x,y,z$, 
\be{tb2}
\begin{array}{lcl}
{{\cal B}}_{v} & = & span \{ \
-i \left( \sigma_{v}\otimes I_{2\times 2} +I_{2\times 2}\otimes 
\sigma_{v} \right) \otimes I_{2\times 2}, \ 
-i (I_{2\times 2}\otimes I_{2\times 2}\otimes \sigma_{v}) \},\\
 {\cal B} & = &  {\cal B}_x \oplus  
{\cal B}_y \oplus  {\cal B}_z.
\end{array}
\ee
To deal with this case, we need to consider three  
subcases:
\bi
\item[(i)]
$|J_{13}|\neq |J_{23}|$,
\item[(ii)]
$J_{13}=J_{23}$,
\item[(iii)]
$J_{13}=-J_{23}$.
\ei
For the case (i) we can apply Theorem \ref{10} and conclude that,
if the associated graph is connected then 
${\cal L}=su(8)$ and the system is
operator controllable. For the case (ii), 
the model will turn out to be neither operator 
controllable nor state controllable. Finally, in the case (iii), the 
controllability properties of the model will depend on the coefficient 
$J_{12}$. In fact the system will be   operator controllable (i.e. 
${\cal L}= su(8)$) if $J_{12}\neq 0$, while, if $J_{12}=0$, then the 
system will be state controllable but not operator controllable (so, 
from Theorem \ref{sommarize}, in 
this case ${\cal L}$ is  isomorphic to $sp(4)$).

\vs

{$\bullet$ {\bf{case (ii): $J_{13}=J_{23}$}}}

{}From a physical point of view, in 
this case the particles one and two feel 
the same magnetic  field and have the same 
interaction with the third 
particle, therefore it is not possible to manipulate separately these 
two 
particles. This internal symmetry 
of the system results in lack of
controllability both for the 
evolution operator and the state. 
If $J_{13}=J_{23}=0$, we have:
\bi
\item
if $J_{12}=0$, then ${\cal L}={{\cal B}}$,
\item
if $J_{12}\neq 0$, then ${\cal L}=span\{A\} \oplus {{\cal B}}$
and the matrix $A$ commutes with all the matrices in $\cal L$.
\ei
Now we consider the case $J_{13}=J_{23}\neq 0$.
We first 
define an 
operation of `symmetrization' $\rho$ on the matrices in $u(4)$, as 
follows:
\be{rho}
i\rho\left( \sigma_{1}\otimes \sigma_{2} \right) = i \frac{1}{2}
\left( \sigma_{1}\otimes \sigma_{2} + \sigma_{2}\otimes 
\sigma_{1}\right),
\ee
with $\sigma_{1},\sigma_{2}\in \{ I_{2\times 
2},\sigma_{x},\sigma_{y},\sigma_{z}\}$, and we extend $\rho$ to all of 
the 
matrices of $u(4)$ by linearity.
Let:
\be{unito}
\rhof =\left\{ X\in u(4) \ | \ \rho(X)=X \right\}.
\ee
Notice that:
\be{impo}
X_{1},X_{2}\in \rhof \Rightarrow X_{1}X_{2}\in \rhof.
\ee
For sake of completeness, we include a proof in Appendix B. Let:
\be{calh}
{\cal H} =  \left \{ H = F \otimes 
\sigma_{j} \ \left | \ \begin{array}{l} \ F\in \rhof, \\
\sigma_{j}\in \{I_{2\times 2}, \sigma_{x}, 
 \sigma_{y},  \sigma_{z}\} \\ H \neq 
i I_{8\times 8} \end{array} \right. \right\}.
 \ee
First, we have:
 \be{elle3}
 {\cal L} \subseteq {\cal H}.
 \ee
To see this, recall that ${\cal L}$ is generated by:
\[
A=-iJ_{12}\left(\sigma_{x}\otimes\sigma_{x}\otimes I_{2\times 2}+
\sigma_{y}\otimes\sigma_{y}\otimes I_{2\times 2}
+\sigma_{z}\otimes\sigma_{z}\otimes I_{2\times 2}\right)
-i J_{13}
\]
\[
\Big(
(\sigma_{x}\otimes I_{2\times 2}+ I_{2\times 2}\otimes 
\sigma_{x})\otimes \sigma_{x} +
(\sigma_{y}\otimes I_{2\times 2}+ I_{2\times 2}\otimes 
\sigma_{y})\otimes \sigma_{y} +
(\sigma_{z}\otimes I_{2\times 2}+ I_{2\times 2}\otimes 
\sigma_{z})\otimes \sigma_{z} \Big),
\]
and by the matrices in ${{\cal B}}$ (see equation (\ref{tb2})).
Thus ${\cal L}\subseteq {\cal H}$ follows {}from the fact that both $A$ 
and ${{\cal B}}$ are in ${\cal H}$, and that ${\cal H}$ is a Lie algebra  
because of  (\ref{impo}).
Now we have:
\bi
\item[(i)]
if $J_{12}\neq 0$, then ${\cal L}={\cal H}$, and it has dimension 39;
\item[(ii)]
if $J_{12}= 0$, then ${\cal L}{\subset}{\cal H}$, where the inclusion 
is strict  
and it has dimension 38.
\ei
The proof of both the previous statements (i) and (ii) follows {}from 
the analysis of the Lie algebra structure for this model, given in the 
Appendix C.
In both cases ${\cal L}$ is not $su(8)$, thus the model is not 
operator controllable. Moreover, by looking at the two possible dimensions
of ${\cal L}$, the model can not be state 
controllable either. In fact to have 
state controllability we would 
need, see Theorem \ref{sommarize},  ${\cal L}=su(8)$ or ${\cal L}$ 
isomorphic to $sp(4)$, which has dimension 36.

\vs
{$\bullet$ {\bf{case (iii): $J_{13}=-J_{23}\not=0$}}}

This case is interesting because it provides a physical example of a 
system which is state controllable but not operator controllable. 
It also shows that for spin 
systems with some  gyromagnetic ratios possibly equal to each other the two 
notions of controllability do not coincide (cfr.  Theorem \ref{9}). 

\vs

Consider the following vector spaces of matrices 

\be{Mm}
{\cal M}:=span \{iI_{1v,3w}-iI_{2v,3w}, \quad v,w\in \{ x,y,z \}\},
\ee
\be{Cc}
\tildec:=span\{iI_{1v}+iI_{2v},iI_{3w}, \quad v,w \in \{x,y,z\}\},
\ee
\be{Nn}
\n:=span  \{ iI_{1v,2w,3p}+iI_{1w,2v,3p}, \quad v,w,p \in \{x,y,z\} 
\},
\ee
\be{Rr}
\r:=span \{ iI_{1v,2w}-iI_{1w,2v}, \quad v \not=w, v,w \in \{x,y,z\}\}.
\ee
It can be seen by verifying the commutation relations among these 
vector spaces that ${\cal A}:=\m \oplus \tildec \oplus \n \oplus \r$ is a 
subalgebra. Moreover, using the test in  part 4 of 
Theorem \ref{sommarize}, it can be 
shown that this Lie algebra is isomorphic to $sp(4)$. It is  
interesting to notice that the decomposition  
${\cal A}:=\m \oplus \tildec \oplus \n \oplus \r$  is underlying a Cartan 
decomposition of $sp(4)$ \cite{Helgason}
since the following inclusions among the above vector spaces hold:
\be{inclusions}
[\tildec \oplus \n,\tildec \oplus \n] \subseteq 
\tildec \oplus \n, \quad [\tildec \oplus \n, \m \oplus \r] 
\subseteq \m \oplus \r, \quad [\m \oplus \r, \m \oplus \r] 
\subseteq \tildec \oplus \n.
\ee
To see that $\cal A$ is a subalgebra of $\cal L$ notice that 
Lemma \ref{6l} gives a basis for 
$\tildec$. By taking the Lie bracket of $A$ with $I_{3x} \in {\cal B}$ and
then of the resulting matrix with $I_{3z} \in {\cal B}$, we obtain a
matrix proportional to $i(I_{1z,3x}-I_{2z,3x})$ and, from this, taking
Lie brackets with elements in $\tildec$ we can obtain all the elements
in the basis of $\cal M$ indicated in (\ref{Mm}). Thus, both $\tildec$ 
and $\m$ are included in ${\cal L}$.
A basis of $\n$ can be obtained by Lie 
brackets of appropriate elements of 
$\m$ (possibly adding an element 
of $\tildec$). Finally, a basis of $\r$ can be obtained by Lie brackets of 
appropriate elements of $\m$ and $\n$. Therefore the Lie algebra 
$\cal A$ is a subalgebra of $\cal L$.   
The two Lie algebras coincide if $J_{12}=0$. 
This is the case remarked above  of 
a system that, according to Theorem \ref{sommarize} is state 
controllable, since $\cal L$ is isomorphic  to $sp(4)$, 
 but not operator controllable. If $J_{12}\not=0$, then 
the matrix $A$ is not in the Lie algebra $\cal A$. However, it is 
still possible to generate $\cal A$, which is isomorphic to $sp(4)$ 
and, applying part 5 of Theorem \ref{sommarize}, we conclude that ${\cal 
L} =su(8)$ in this case, and the system is operator controllable.  

\vs

The results of this section and Section 
\ref{SPEGR} remain true even if
we set $u_z= constant$ in the model (\ref{scrod}) if we assume that
there exist no two values for the gyromagnetic ratios $\gamma_1$ and
$\gamma_2$ such that $\gamma_1 = -\gamma_2$ (cfr. Remark \ref{RDM})

%same considerations of Remark \ref{RDM} hold if the control 
%$u_{z}$ is held constant, showing that the above analysis remains 
%unchanged in this case. 

\section{Conclusions}
\label{c}

We have presented an analysis of the Lie algebra structure 
associated to a system of $n$ spin $\frac{1}{2}$ particles with
different gyromagnetic ratios and inferred
its controllability properties. These only depend on the properties of a
graph obtained by connecting two nodes representing two 
particles if the coupling
constant between the two particles is different {}from
zero. Controllability of the state and of the unitary evolution operator
are equivalent for this class of systems. If the system is not
controllable then it is a parallel connection of a number of
controllable systems equal to the connected components of the
associated graph. The latter result can be easily generalized to the case
where the connected components do not represent controllable subsystems,
which is a case that might occur if some of the gyromagnetic ratios are equal. 

We have given a complete description of the low 
dimensional cases (up to a number of particles equal to three) with 
isotropic interaction and possibly equal gyromagnetic ratios. This 
analysis  is of  interest since, in many physical 
situations,  
a small number of particles is controlled. These  results also 
provide an example of a quantum system which is controllable in the 
state but not in its unitary evolution operator. Thus, the equivalence 
of the two notions of controllability, proved for spin systems in  the case of 
different gyromagnetic ratios, is  no longer true if some of the 
gyromagnetic ratios are equal. 

This paper also presented a general sufficient condition of
controllability for spin systems in terms of the associated graph.

\section*{Appendix  A: Some properties of
the matrices $I_{k_1 l_1, k_2l_2 ,..., k_rl_r}$}

We collect in this appendix some properties of the matrices
$I_{k_1 l_1, k_2l_2 ,..., k_rl_r}$, in particular involving the
commutators of these matrices. These
relations can be easily proven by
using the fundamental property:
\be{fundapro}
[A \otimes B, C \otimes D]=[A,C] \otimes BD +CA \otimes [B,D],
\ee
where $A$ and $B$ are square
matrices of the same dimensions and $B$ and
$D$ are square matrices of the same dimensions as well.

\vs

\noindent {\bf Property 1:} If
$\{k_1,k_2,...,k_r\} \cap  \{\bar k_1, \bar k_2,...,\bar k_s\} =
\emptyset$ then
\be{Prop1}
[I_{k_1l_1,k_2l_2,...,k_rl_r}, I_{\bar k_1m_1,\bar k_2m_2,...,\bar
k_sm_s}]=0,
\ee
for every possible combination of $l_j$'s and $m_j$'s.

\vs

\noindent {\bf Property 2:} Assume that 
$\bar k \in \{k_1, k_2, ..., k_r \}$, and, in
particular, $\bar k=k_j$.
\begin{itemize}
\item[(a)]
 If $l_j=m$, then
\be{Prop2}
[I_{k_1l_1,k_2l_2,...,k_rl_r}, I_{\bar k,m}]= 0,
\ee
\item[(b)]
if $[\sigma_{l_j},\sigma_m]=\alpha \sigma_{\tau}$, with $\alpha=\pm i$,
then:
\be{Prop2bis}
[I_{k_1l_1,k_2l_2,...,k_rl_r}, I_{\bar k,m}]\, =\,
\alpha \,
I_{k_1l_1,\ldots,k_j\tau,\ldots,k_rl_r}.
\ee
\end{itemize}

\section*{Appendix B: Proof of property (\ref{impo})}

In order to see (\ref{impo}), we write any element 
$X$ of $u(4)$ as follows (we 
use the definition $\sigma_{0}:=I_{2 \times 2}$ and the ordering 
$0 < x <y<z$)
\[
X= \sum_{j,k=0,x,y,z}\alpha_{jk}i \sigma_{j}\otimes \sigma_{k}=
\sum_{j=0,x,y,z}\alpha_{jj}i \sigma_{j}\otimes \sigma_{j} + 
\]
\[
+
\sum_{j<k} \frac{\alpha_{jk}+\alpha_{{kj}}}{2}(\sigma_{j} \otimes 
\sigma_{k}+\sigma_{k}\otimes \sigma_{j})+
\sum_{j<k}\frac{\alpha_{{jk}}-\alpha_{{kj}}}{2}(-\sigma_{k}\otimes 
\sigma_{j}+\sigma_{j}\otimes \sigma_{k}).
\]
{}From this expression, it is immediate to see that 
$X\in \rhof$ if and only the terms in the last sum are all zero. Therefore 
a basis of $\rhof$ is given by the matrices of the form
\be{base}
i\left( \sigma_{l}\otimes \sigma_{v}+\sigma_{v}\otimes 
\sigma_{l}\right),
\ee
with $l,v=0,x,y,z$.
In view of this fact, it is sufficient to verify (\ref{impo}) on all the 
matrices of the form (\ref{base}). 
This last fact is only a straightforward calculation.

\section*{Appendix C: Structure of 
the Lie algebra $\cal L$ in the
case ${n=3}$, $J_{13}=J_{23}\neq 0$ 
$(\gamma_{1}=\gamma_{2}\not=\gamma_{3})$}

We look at the following vector 
subspaces of $\cal H$ ($\cal H$ 
is the vector space defined in 
(\ref{calh}).

\begin{eqnarray}
\tilde {\cal C}:=span \quad i\{ I_{1v}+I_{2v},I_{3w} \quad v,w=x,y,z \} \\
{\cal M}:=span \quad i \{ I_{1v,3w}+I_{2v,3w} \quad v,w=x,y,z \} \\
{\cal N}:=span \quad i \{I_{1v,2w,3p}+I_{1w,2v,3p} \quad v,p,w=x,y,z
\}\\
{\cal Q}:= span \quad i \{ I_{1v,2w}+I_{1w,2v} \quad v \not=w=x,y,z\}\\
{\cal R}:= span \quad i \{I_{1x,2x}-I_{1y,2y},I_{1x,2x}-I_{1z,2z}\}
\end{eqnarray}

The following commutation
relations are easily verified
\be{CCCC}
[\tilde {\cal C}, \tilde {\cal C}] \subseteq \tilde {\cal C},
\quad [\tildec,\m] \subseteq \m, \quad [\tildec,\n] \subseteq \n,
 \quad
[\tildec,\q] \subseteq \n \oplus \r, \quad  [\tildec,\r] \subseteq \q;
\ee
\be{MMMM}
[\m,\m] \subseteq \tildec \oplus \m \oplus \n, \quad
[\m,\n] \subseteq \m\oplus \q \oplus \r, \quad
[\m,\q] \subseteq \n \quad
[\m,\r] \subseteq \n;
\ee
\be{NNNN}
[\n,\n] \subseteq \tildec \oplus \n, \quad
[\n,\q] \subseteq \tildec \oplus \m, \quad
[\n,\r] \subseteq \m;
\ee
\be{QQQQ}
[\q,\q] \subseteq \tildec, \quad
[\q,\r] \subseteq \tildec;
\ee
\be{RRRR}
[\r,\r]=0.
\ee

A basis of  $\tilde {\cal C}$ is generated by the matrices
$B_{x}$, $B_{y}$, $B_{z}$ according to Lemma \ref{6l}, while
a basis in $\cal M$ can be obtained
calculating the Lie bracket of  $A$ with $I_{3x} \in {\cal C}$ then
taking the Lie bracket with $I_{3z}$ so as to obtain
$I_{1z,3x}+I_{2z,3x}$. From this, taking the Lie bracket with elements
in $\cal C$, we can obtain all the elements in the basis of $\cal M$ in
(\ref{MMMM}). A basis of  $\n$
is obtained by Lie brackets of elements in $\m$. In particular,
to obtain elements of the form $iI_{1v,2v,3w}$, we calculate
$[iI_{1v,3l}+iI_{2v,3l},-iI_{1v,3p}-iI_{2v,3p}]-\frac{1}{2}iI_{3w}$,
with $p\not=l$ $v,p,l \in \{x,y,z\}$ and 
$i\sigma_{w}=[\sigma_{l},\sigma_{p}]$. To
obtain elements of the form $iI_{1v,2w,3p}+iI_{1w,2v,3p}$,
$v\not=w$, $v,w,p \in \{x,y,z\}$, we can calculate Lie brackets of
elements of the form $iI_{1m,3m}+iI_{2m,3m}$, with elements of the
form $iI_{1n,3n}+iI_{2n,3n}$, with $n \not=m$ and, possibly, calculate
the Lie bracket with an element of the form $iI_{3l}$, $l,m,n \in
\{x,y,z\}$. A basis of $\q$ can be obtained by Lie brackets
between elements of the form $iI_{1v,2v,3x} \in \n$ with elements of
the form $iI_{1w,3x}+iI_{2w,3x} \in \m$, $v\not=w,$ $v,w \in
\{x,y,z\}$. A basis  of $\r$ can  be obtained by the Lie
bracket of elements $iI_{1v,2w,3x}+iI_{1w,2v,3x} \in \n$
with elements $iI_{1p,3x}+iI_{2p,3x}$, with $p\not=v\not=w$, $p,v,w
\in \{x,y,z\}$.

Notice that if $J_{12}=0$, then $A$ is an element of the Lie algebra
above described $\tildec \oplus \m \oplus \n \oplus \q \oplus \r$, while if
$J_{12} \not=0$ we have $\tildec \oplus \m \oplus \n \oplus \q \oplus
\r={\cal H}/span\{A\}$, and the Lie algebra $\cal L$, in this case,
coincides with $\cal H$.

\end{document}